# The elastic stiffness tensor of cellulosic viscose fibers measured with Brillouin spectroscopy

Caterina Czibula[1,2,*], Manfred H. Ulz[3], Alexander Wagner[1], Kareem Elsayad[4], Ulrich Hirn[1], Kristie J. Koski[2]

[1] Institute of Bioproducts and Paper Technology, Graz University of Technology, Inffeldgasse 23, 8010 Graz, Austria

[2] Department of Chemistry, University of California Davis, 1 Shields Ave. 222 Chemistry, Davis CA, 95616, USA

[3] Institute of Strength of Materials, Graz University of Technology, Kopernikusgasse 24, 8010 Graz, Austria

[4] Division of Anatomy, Center for Anatomy and Cell Biology, Medical University of Vienna, Währinger Straße 13, 1090 Vienna, Austria

* corresponding author: caterina.czibula@tugraz.at

**Abstract**

Brillouin light scattering spectroscopy (BLS) is applied to study the micromechanics of cellulosic viscose fibers, one of the commercially most important, man-made biobased fibers. Using an equal angle scattering geometry, we provide a thorough description of the procedure to determine the complete transversely isotropic elastic stiffness tensor. From the stiffness tensor the engineering-relevant material parameters such as Young's moduli, shear moduli, and Poisson's ratios in radial and axial fiber direction are evaluated. The investigated fiber type shows that, at ideal conditions, the material exhibits optical waveguide properties resulting in spontaneous Brillouin backscattering which can be used to obtain additional information from the Brillouin spectra, enabling the measurement of two different scattering processes and directions with only one scattering geometry.

1. Introduction

Viscose fibers are man-made fibers from the cellulose of wood pulp. Being biobased and fully biocompatible, viscose is one of the most widely used materials for clothing, upholstery, textiles, hygiene products, as well as food applications [1]. Cellulose, from which viscose is made, is the most abundant biopolymer and primary component of plant-based materials. In plants, cellulose forms complex hierarchical fibers with other biopolymers like hemicelluloses, pectins and lignins. Like most fibrous materials in biological matter, cellulosic fibers exhibit directionality in their mechanical properties. These fibers can consist of several cell wall layers which have varying chemical composition, thickness, and cellulose fibril orientation [2,3]. In this work, we consider man-made cellulose fibers formed during the so-called viscose process. During the viscose process, wood cellulose is dissolved in a solvent and subsequently spun into a yarn. Thus, viscose fibers differ from wood fibers by consisting solely of cellulose which has a different crystalline structure in man-made fibers (cellulose II) and is lacking the hierarchical structural complexity of the wood fibers [4]. Moreover, their cross-section is tunable during the production process. Viscose fibers exhibit many similarities to natural fibers, but are less complex in structure and chemistry. Therefore, they serve as a suitable simplified model system for understanding the more complex mechanical properties of natural cellulosic fibers [5,6]. Due to their industrial relevance, understanding their mechanical properties is important for realizing desirable mechanical properties of the end products, though few investigations of elastic properties of viscose are available in academic literature.

Fibrous materials like viscose fibers have directional – anisotropic – mechanical properties. In the axial direction – which is along the long axis of the fiber – mechanical properties such as stiffness are usually higher than in radial direction of the fiber – perpendicular to the long axis of the fiber. For such materials, it is not sufficient to measure the material only in one direction. One needs to obtain experimental data in multiple directions in order to determine the full elastic stiffness tensor and to understand how the fiber behaves under mechanical load. Mechanical modeling of biobased fibrous materials – e.g., in network materials [7] or composite materials [8,9] has become highly relevant. For these models the complex, directional mechanical behavior of the fibers has to be determined experimentally. However, the testing of fibrous materials is challenging with classical mechanical testing methods because the fibers are only a few millimeters in length and $10 - 50\,\mu m$ in width, which explains the dearth of elastic measurements of viscose fibers. The associated experimental equipment for these small objects is usually custom built, elaborate, and difficult to handle [10–13].

Brillouin light scattering spectroscopy (BLS) is the inelastic interaction of laser light with acoustic phonons in a material [14]. This non-contact method does not require complicated sample preparation procedures and is not affected by the testing method, which can introduce errors. In recent years, BLS has been commonly applied in biological and biomedical studies. With the assumption of isotropy – the mechanical behavior being independent of the material direction – different soft biological matter was studied [15–20]. Further, BLS can measure all the elastic stiffness tensor elements and, thus, provide the complete mechanical properties of a material. BLS has proven to be a suitable approach to determine the full elastic stiffness tensor of biological fibers such as collagen [21], fibrous proteins [22], spider silk [23], bamboo [24] and marine sponges [25]. Previously, the anisotropy of plant fibers was investigated [26] and we have measured the transverse modulus of viscose fibers with BLS [27].

In this work, we measure the directional elastic stiffness and mechanical properties of a viscose fiber and demonstrate that there is a clear difference between the stiffness of a material, which is described by the elastic stiffness tensor directly measured with BLS, and the engineering material parameters such as Young's modulus, which is obtained by mechanical testing. We use a comprehensive experimental procedure which can be applied to any transversely isotropic fibrous material to obtain the full elastic stiffness tensor by BLS. We use an equal angle (90a) scattering geometry that has substantial potential in BLS. This geometry neglects the refractive index of the material, which is difficult to determine for heterogeneous materials. Finally, we compare the measured elastic stiffness tensor to the engineering mechanical parameters from mechanical testing (Young's moduli, Poisson's ratios and shear moduli).

## 2. Materials & Methods

### 2.1 Fiber samples and preparation

The samples used in this work were industrial flat viscose fibers (9 dtex, 150 µm width) with a rectangular cross-section [5,27] as presented in the optical micrograph in Fig. 1a. The fibers were attached to metallic sample holders with a 1 cm hole in the middle and fixed on both sides with double-sided tape. Care was taken to ensure that the fiber was straight across the gap. For the measurements, the center of the fiber was chosen in areas where the fiber was not tilting to a high extent (they have a natural curvature and kinks due to the production process). Based on literature data, the density of the material is assumed as $1500 \; kg/m^3$ [28].

### 2.2 Single fiber tensile testing

Ten single viscose fiber samples were tested with a Dynamic Mechanical Analyzer (DMA800, TA Instruments, USA) with a preload of $3 \; mN$. The tests were performed displacement-controlled with a displacement rate of 10 µm/s. The cross-sectional area of $(620 \pm 60)$ µm$^2$ was determined from three individual viscose fibers by a procedure based on microtome-cutting [13,29,30], see Figure 1b.

### 2.3 Brillouin light scattering setup

Brillouin scattering data were acquired with a six-pass tandem, scanning Fabry–Perot interferometer (TFP-1, The Table Stable Ltd., Switzerland) in 90a scattering geometry. The mirror spacing, which controls the scanned frequency range, was fixed to 4.5 mm with a scan rate of 500 nm to achieve a free spectral range of about ±30 GHz. A pinhole with a diameter of 450 microns was used to achieve a finesse of approximately 100-110. An experimental optics setup with a confocal arrangement was used to focus and collect light from the sample. A green laser beam (λ = 532.15 nm) with a power of <10 mW was focused onto the samples from a neodymium yttrium vanadate laser (Coherent Verdi V6). The focusing of the light onto the sample was provided by a 10X objective (focusing the incident light, Obj 1 in Fig. 2a, Mitutoyo M Plan Apo SL, NA = 0.28, f = 200 mm) and 20X objective (focusing the light scattered from the sample, Obj 2 in Fig. 2a, Mitutoyo M Plan Apo SL, NA = 0.28, f = 200 mm). A small numerical aperture on the collection objective was used to limit the collection angle in the sample for highest possible resolution.

The fiber was suspended on a 25 mm stainless steel gasket with a 5-7 mm hole and fixed into an optic rotation mount (Newport RSP-1T). Input polarization of the laser was rotated using a λ/2 plate with pre-selection and post-selection of the incident and scattered light using a linear polarizer with an extinction ratio >10,000:1. Spectra were collected with acquisition times of 4-12 h. The acquisition time was influenced by the sample itself, the position on the sample, and the scattering geometry. The relative humidity (RH) varied between 30 % and 50 % RH whereas the room temperature was maintained at 20-21°C.

**2.4 Theory of elastic waves in crystals**

The treatment of elastic waves in crystals is well established in the literature [31,32] and has been applied manyfold [21–25,33–35]. Here, we summarize a route to obtain the full elastic stiffness tensor for a broader community with a more elaborate introduction. For the reader, who is just interested in the main information, we suggest to skip to the paragraph before Eq. 3 is stated. In the following, we give a brief overview by considering infinitesimal strain theory and introducing the (Cauchy) stress tensor $\sigma$ and the (engineering) strain tensor $\varepsilon$. In the three-dimensional case (for i, j, k, l = x, y, z or 1, 2, 3), Hooke's law is written as:

$$\sigma_{ij} = C_{ijkl}\varepsilon_{kl}, \qquad \text{(Equation 1)}$$

the tensor of fourth-order $C_{ijkl}$ denotes the elastic stiffness tensor. As a fourth-order tensor, $C_{ijkl}$ could have $3^4 = 81$ independent entries. However, $C_{ijkl}$ of any crystal system has at least three symmetry properties, which ultimately limit the number of independent entries to 21: (i) the strain tensor $\varepsilon_{kl}$ is symmetric and has 6 independent components instead of 9, and, (ii) the stress tensor $\sigma_{ij}$ is symmetric as well. The first and the second symmetry properties lead to the minor symmetry property, $C_{ijkl} = C_{jikl} = C_{ijlk} = C_{jikl}$, and reduce $C_{ijkl}$ to 36 independent entries. (iii) We assume, as is standard procedure in solid state mechanics, the existence of a strain energy potential $\psi(\varepsilon)$ and require that:

$$C_{ijkl} := \frac{\partial^2 \psi(\varepsilon)}{\partial \varepsilon_{ij} \partial \varepsilon_{kl}}. \qquad \text{(Equation 2)}$$

This reveals that there must be symmetry in the indices $ij$ and $kl$ as we derive a scalar potential twice by the same tensor. As a result, the major symmetry property, $C_{ijkl} = C_{klij}$, is revealed and the independent entries of $C_{ijkl}$ reduce to 21.

A fiber is considered radially symmetric around its axis, with anisotropy from radial to axial. A good model that matches this symmetry is the hexagonal crystal system, which is isotropic in the a-axis (similar to the radii of a fiber) and has anisotropy from the a-axis to the elongated c-axis, see Fig. 1c.

Hexagonal symmetry is often applied in the literature to fibers [21,36,37] given these have similar symmetry properties. This symmetry reduces the number of independent entries in the elastic stiffness tensor from 21 to only five. Therefore, the elastic stiffness tensor can be written by switching from indicial notation to so-called Voigt's notation:

$$C_{\alpha\beta} = \begin{bmatrix} C_{11} & C_{12} & C_{13} & 0 & 0 & 0 \\ C_{12} & C_{11} & C_{13} & 0 & 0 & 0 \\ C_{13} & C_{13} & C_{33} & 0 & 0 & 0 \\ 0 & 0 & 0 & C_{44} & 0 & 0 \\ 0 & 0 & 0 & 0 & C_{44} & 0 \\ 0 & 0 & 0 & 0 & 0 & C_{66} \end{bmatrix},$$ (Equation 3)

$$\text{with } C_{12} = C_{11} - 2C_{66}.$$ (Equation 4)

A hexagonal crystal system belongs to the class of transversely isotropic materials. A transversely isotropic material has a preferred direction and is isotropic in the plane perpendicular to this direction. In Eq. 3 and Fig. 1c the x, y-plane is the plane of isotropy, while the z-direction is the preferred material direction. Most fibers can be classified as transversely isotropic materials.

We note here that the introduced strain energy potential $\psi(\boldsymbol{\varepsilon})$ (compare Eq. 2) is commonly assumed to be zero at zero strain giving the reference state. Any deformation must increase the strain energy potential and, consequently, the strain energy potential cannot be smaller than zero. This requires the components of the elastic stiffness tensor of a hexagonal crystal system to satisfy the following three constraints: $C_{11} > |C_{12}|$, $(C_{11} + C_{12})C_{33} > 2C_{13}^2$, $C_{44} > 0$. If these constraints are not satisfied, the crystal system cannot be considered stable.

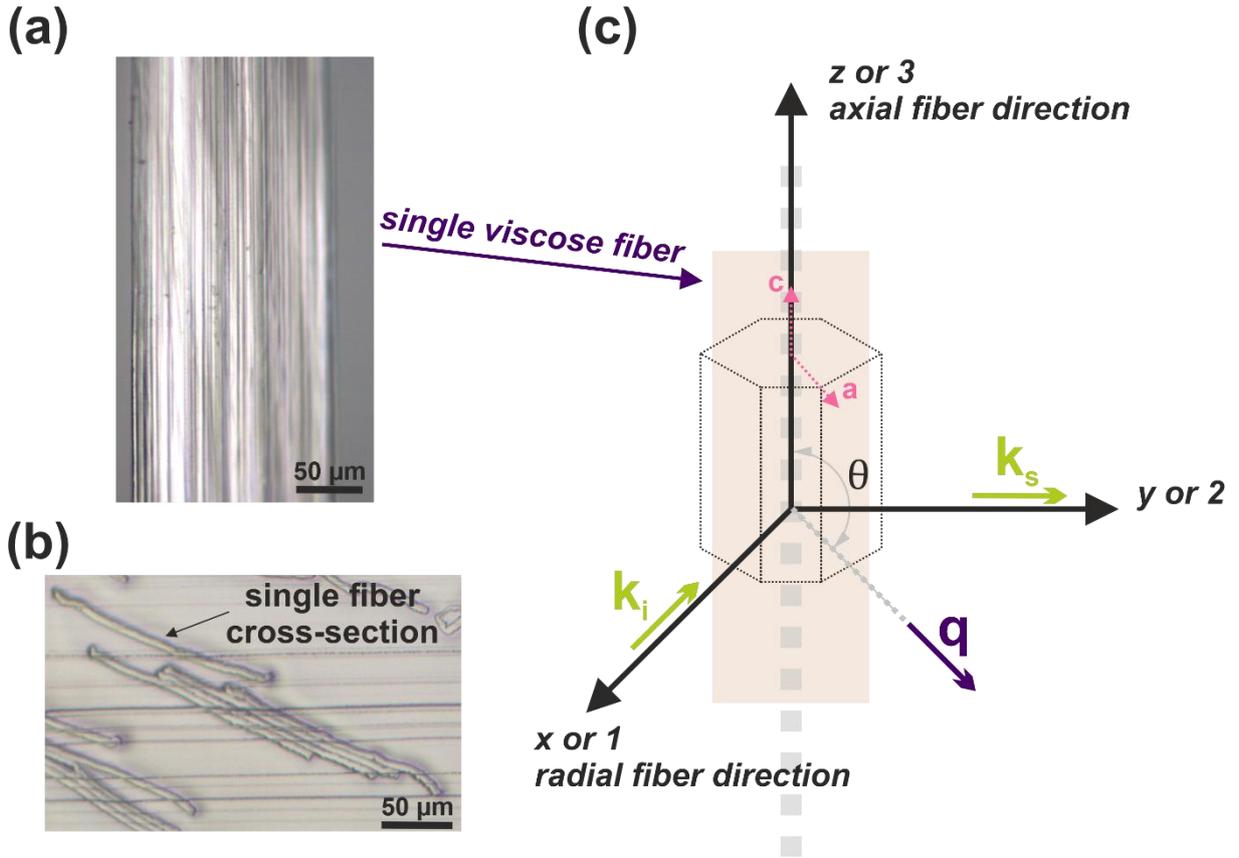

*Figure 1: Optical image of (a) the investigated viscose fiber type and of (b) typical cross-sections of the investigated grade of viscose fibers. (c) Sketch of the viscose fiber and its relation to the hexagonal crystal structure. A regular hexagon is presented with the a- and c-axes to depict the sixfold rotation symmetry about the z-axis. The wave normal **q** has the angle θ to the z-axis. The incoming and scattered light are depicted by **$k_i$** and **$k_s$**, respectively. The axes notation is given as either 1, 2, and 3 or x, y, and z since both are commonly applied in the literature.*

This sets the mechanical basis for the approach of determining the elastic stiffness tensor, however, now the connection to the elastic waves which are travelling through the fibrous material needs to be established. This will be done by solving the Christoffel equation for the hexagonal symmetry case which we assume for fibrous materials.

From d'Alembert's principle, we can derive the equation of motion of a deformed body. In the absence of gravity forces, we can state it as: $\rho \ddot{u}_i = \partial \sigma_{ij} / \partial x_j$, with ρ being the density of the medium. The displacement vector field $\mathbf{u}(x,t)$ is introduced in this equation, which can be thought of as the displacement of particles dependent on position $x$ and time $t$. With the (linearized) strain $\varepsilon_{kl} = \frac{1}{2}(\partial u_k/\partial x_l + \partial u_l/\partial x_k)$ and Eq. 1 we obtain the following linear homogeneous second-order differential equation in the displacement vector

$$\rho \frac{\partial^2 u_i}{\partial t^2} = C_{ijkl} \frac{\partial^2 u_l}{\partial x_j \partial x_k} \ .$$

(Equation 5).

We focus on plane monochromatic waves, which satisfy the above differential equation. The displacement vector of such a wave can be written as $\boldsymbol{u} = \boldsymbol{u}^0 e^{i\phi}$ with $\phi = \boldsymbol{k} \cdot \boldsymbol{x} - \omega t$. Here, we have introduced the vector amplitude $\boldsymbol{u}^0$, the angular frequency $\omega$ and the wave vector $\boldsymbol{k} = k\boldsymbol{q}$, with $\boldsymbol{q}$ being the wave unit normal vector. Clearly, only certain values of $\boldsymbol{u}^0$, $\omega$ and $\boldsymbol{k}$ will satisfy Eq. 5. As a result, we obtain with the sound velocity $V = \omega/k$ of the wave and the Kronecker delta $\boldsymbol{\delta}$, the Christoffel's equation:

$$\left(\frac{1}{\rho} C_{ijkl} q_j q_k - V^2 \delta_{il}\right) u_l = 0_i. \tag{Equation 6}$$

The Christoffel's equation is important to the theory of elastic waves in crystals and demonstrates that the problem reduces to solving for the eigenvalues $v$ and eigenvectors $\boldsymbol{u}$ of the symmetric second-rank tensor $\Lambda = \frac{1}{\rho} C_{ijkl} q_j q_k$.

In the general case, there are three distinct eigenvalues, each having an eigenvector defining the direction of the displacement in the wave. Therefore, for any given $\boldsymbol{q}$ there are three perpendicular waves with different eigenvectors $\boldsymbol{u}$ and sound velocities $V$. The wave with the eigenvector $\boldsymbol{u}$ that is closest to the wave normal $\boldsymbol{q}$ is called the quasi-longitudinal wave. This wave will mainly vibrate along the direction of propagation but will show a small portion of vibration perpendicular too. If the wave's eigenvector $\boldsymbol{u}$ coincides with the wave normal $\boldsymbol{q}$, then this wave is called a purely longitudinal wave and will only vibrate along the direction of propagation. This quasi-longitudinal (L) wave will have the largest eigenvalue and therefore denotes the fastest wave. The waves associated with the other two eigenvectors are called either quasi-transverse or purely transverse waves, depending on if they present oscillations close or completely perpendicular to the direction of propagation. These waves are further divided into the fast (T1) and the slow (T2) (quasi-)transverse wave, according to the magnitude of their eigenvalues. It is noteworthy that in most crystal systems, the three velocity curves may only touch each other in certain places, but they do not intersect in the wave vector – frequency dispersion plots. However, hexagonal crystals are an exception to this rule. Here, it is under certain circumstances possible that the transverse waves can cross over each other, but only at one crossing point [38].

Finally, there is a distinct possibility that the two eigenvalues of the transverse waves may coincide, resulting in the degeneration of these two waves into one. This degeneration happens for isotropic materials too, e.g., amorphous materials including glasses.

A hexagonal crystal possesses the elastic stiffness tensor as presented in Eq. 3. If the crystal is aligned to a coordinate system as shown in Fig. 1c and the wave normal has the angle $\theta$ to the sixfold z-axis, then the following analytical equations for the three eigenvalues can be obtained by solving the eigenvalue problem given by the Christoffel's equation in Eq. 6 and involving some algebra [21]:

$$2\rho V_L^2(\Theta) = C_{11} \sin^2(\Theta) + C_{33} \cos^2(\Theta) + C_{44} + \{[(C_{11} - C_{44})\sin^2(\Theta) + (C_{44} - C_{33})\cos^2(\theta)]^2 + 4(C_{13} + C_{44})^2 \sin^2(\theta)\cos^2(\theta)\}^{\frac{1}{2}}$$

(Equation 7)

$$2\rho V_{T1}^2(\Theta) = C_{11} \sin^2(\Theta) + C_{33} \cos^2(\Theta) + C_{44} - \{[(C_{11} - C_{44})\sin^2(\Theta) + (C_{44} - C_{33})\cos^2(\theta)]^2 + 4(C_{13} + C_{44})^2 \sin^2(\theta)\cos^2(\theta)\}^{\frac{1}{2}}$$

(Equation 8)

$$\rho V_{T2}^2(\Theta) = C_{66} \sin^2(\Theta) + C_{44} \cos^2(\Theta)$$

(Equation 9)

## 2.5 Experimental procedure to obtain the elastic stiffness tensor

In this work, we employed solely the 90a-scattering geometry (a top view of schematic is presented in Fig. 2a). The sample is orientated 45° to the incoming and scattered light beam. Furthermore, the sample is hit by the incoming light from the backside of the fiber, and the transmitted scattered light, which is collected from the area in focus, is exclusively captured by the spectrometer. By using the 90a scattering geometry, the sound velocity $V_{90a}$ (Eq. 10), is only dependent on the frequency shift $\Delta f$ and the laser wavelength $\lambda_0$ $(= 532\ nm)$. This is advantageous for heterogeneous, biological materials since no information on the refractive index is required. The frequency shift is determined from the Brillouin peaks via Lorentzian fitting and by employing Eq. 10 the sound velocity $V_{90a}$ can be easily calculated.

$$V_{90a} = \frac{\Delta f \lambda_0}{\sqrt{2}}$$

(Equation 10)

As illustrated in Fig. 2b, the wave normal $\boldsymbol{q}$ is pointing in the 90a direction at an angle of 45° and by rotating the fiber in this geometry different directions within the fiber can be probed. This is illustrated in Fig. 2c. The wave normal $\boldsymbol{q}$ direction is fixed due to the 90a scattering geometry and the only change made to access different fiber directions is by an in-plane rotation of the fiber from its axial direction (0°) towards its radial direction (90°). This is depicted in the side view of the 90a setup in Fig. 2d. Here, the fiber is mounted in 90° orientation and the mount allows a full in-plane rotation of the fiber.

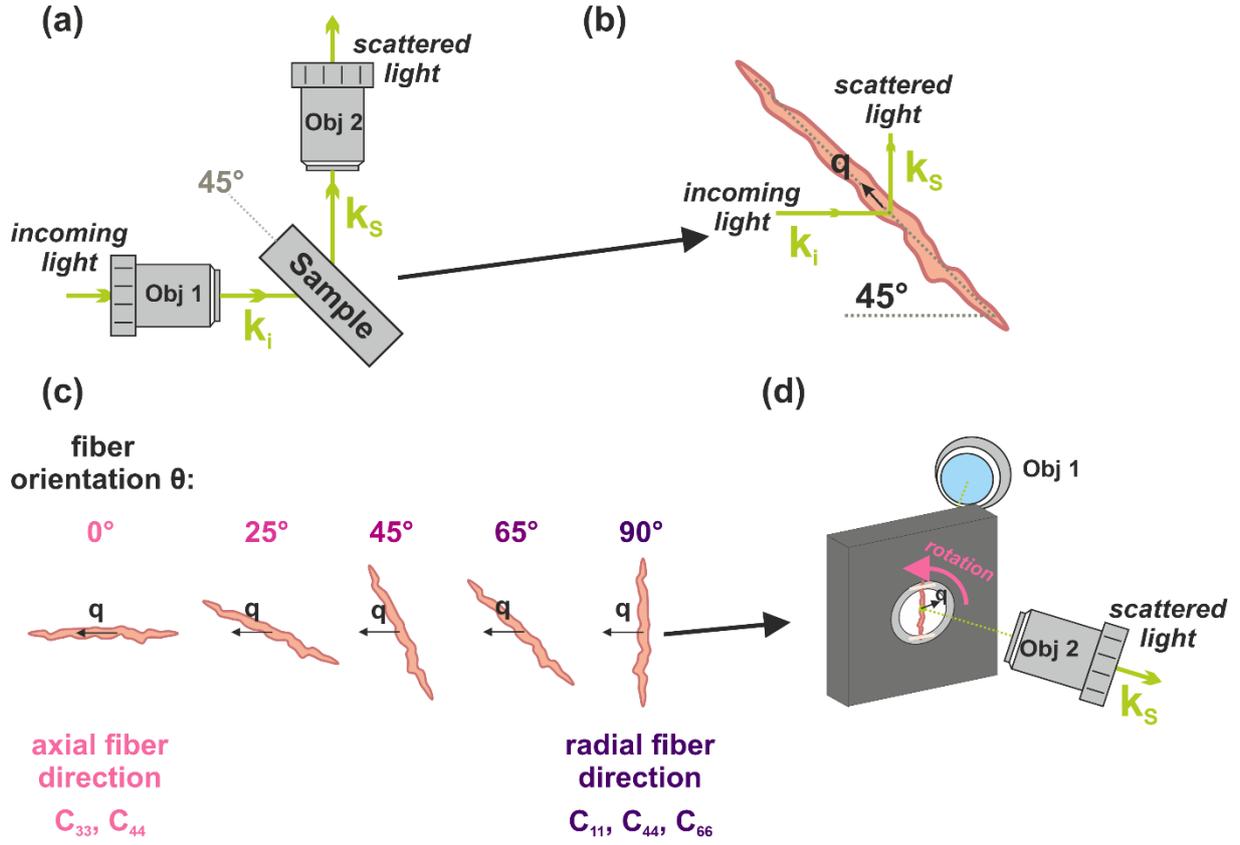

*Figure 2: (a) Sketch of the 90a scattering geometry aligned in top view (Obj 1 – Objective 1, Obj 2 – Objective 2) indicating also the incoming $k_i$ and scattered $k_s$ light, (b) top view sketch of the fiber alignment in relation to $k_i$ and $k_s$ as well as the direction of the wave normal $q$. (c) Rotation of the fiber between 0° (axial fiber direction) and 90° (radial fiber direction) within the 90a scattering geometry in relation to the wave normal $q$. It is possible to obtain all the diagonal components of the hexagonal elastic stiffness tensor ($C_{11}$, $C_{33}$, $C_{44}$, $C_{66}$) from the axial and radial fiber directions. (d) Side view of the 90a scattering geometry with the fiber mounted in 90° orientation (radially) to better visualize the rotation direction of the fiber during the experiments.*

To be able to determine the full elastic stiffness tensor of a hexagonal system, each component of the diagonal of the elastic stiffness tensor $C_{11}$, $C_{33}$, $C_{44}$, and $C_{66}$ – as depicted in Eq. 3 – needs to be directly measured by BLS. BLS measurements at different fiber orientations are necessary to obtain the different components of the elastic stiffness tensor. Since it consists of five independent components – $C_{11}, C_{33}, C_{44}, C_{66}, C_{13}$ – sufficient measurements need to be obtained (i) to determine all the diagonal components, and (ii) to have enough fitting points for the Christoffel solutions. ($C_{12}$ is obtained by Eq. 4 which is dependent on $C_{11}$ and $C_{66}$.)

Here, it is noteworthy to explicitly state two special cases of the given wave normal which are also illustrated in Fig. 2c. First, the fiber orientation $\theta = 0°$ is equal to a coinciding wave normal $q$ with the z-axis, which is corresponding to the axial fiber direction. Next to the quasi-longitudinal wave, the two quasi-transverse waves degenerate in this case since the two transverse velocities coincide. Therefore, this fiber orientation enables the determination of two diagonal components $C_{33}$ and $C_{44}$ of the elastic stiffness tensor:

$$V_L^2 = \frac{C_{33}}{\rho}, \quad V_{T1}^2 = V_{T2}^2 = \frac{C_{44}}{\rho}. \qquad \text{(Equation 11)}$$

Second, the fiber orientation θ=90° – corresponding to the radial fiber direction – is equal to a perpendicular wave normal to the z-axis. This orientation gives – ideally – three diagonal components $C_{11}$, $C_{44}$ and $C_{66}$ of the elastic stiffness tensor:

$$V_L^2 = \frac{C_{11}}{\rho}, \quad V_{T1}^2 = \frac{C_{44}}{\rho}, \quad V_{T2}^2 = \frac{C_{66}}{\rho}. \qquad \text{(Equation 12)}$$

As a fitting strategy, we propose to first obtain the quasi-longitudinal and degenerated quasi-transverse waves for θ=0° (axial fiber direction). With this, the components $C_{33}$ and $C_{44}$ can be determined according to Eq. 11. Then, by acquiring the quasi-longitudinal wave for θ=90° (radial fiber direction), we determine $C_{11}$ directly with Eq. 12. Furthermore, with the quasi-longitudinal mode at some angle θ the constant $C_{13}$ can be obtained with Eq. 7. Ideally, the last constant, $C_{66}$, will be obtained by Eq. 12 as well, however, the slow quasi-transverse mode $v_{T2}$ needs to be known for $C_{66}$ and this constant appears in Eq. 9 only. Eq. 8 can be used to verify the fitting results and determine if the resulting $v_{T1}$ matches with the experiments.

The elastic stiffness components of a transversely isotropic material are related to the classical engineering mechanical properties – Young's moduli $E_{axial}$ and $E_{radial}$, shear moduli $G_{13}$ and $G_{12}$, and Poisson's ratios $v_{13}$ and $v_{12}$ in the following way and in correspondence with the directions indicated in Fig. 1c [23,25]:

$$E_{axial} = E_L = E_z = C_{33} - \frac{2C_{13}^2}{C_{11}+C_{12}} \qquad \text{(Equation 13)}$$

$$E_{radial} = E_T = E_x = E_y = \frac{(C_{11}-C_{12})(C_{33}(C_{11}+C_{12})-2C_{13}^2)}{C_{11}C_{33}-C_{13}^2} \qquad \text{(Equation 14)}$$

$$G_{13} = G_{LT} = G_{zy} = G_{zx} = C_{44} \qquad \text{(Equation 15)}$$

$$G_{12} = G_{TT} = G_{xy} = C_{66} \qquad \text{(Equation 16)}$$

$$v_{13} = v_{LT} = v_{zy} = v_{zx} = \frac{C_{13}}{C_{11}+C_{12}} \qquad \text{(Equation 17)}$$

$$v_{12} = v_{TT} = v_{xy} = \frac{C_{12}C_{33}-C_{13}^2}{C_{11}C_{33}-C_{13}^2} \qquad \text{(Equation 18)}$$

It should be noted that in Eq. 13 – 18, all possible notations mentioned in literature which also correspond to the coordinate system indicated in Fig. 1c are listed for completeness. Mechanical properties with the index T govern the radial fiber direction, while properties with the index L govern the axial fiber direction.

It is important to note that if either elastic stiffness tensor components or mechanical parameters are known, one can obtain the other expressions via Eq. 13 – 18.

In Figure 3, all the basic experimental and theoretical steps to obtain the elastic stiffness tensor of a fibrous material are summarized (under the assumption that it is possible to obtain all the necessary Brillouin peaks for the diagonal components of the elastic stiffness tensor in Eq. 3 from a minimum number of experiments).

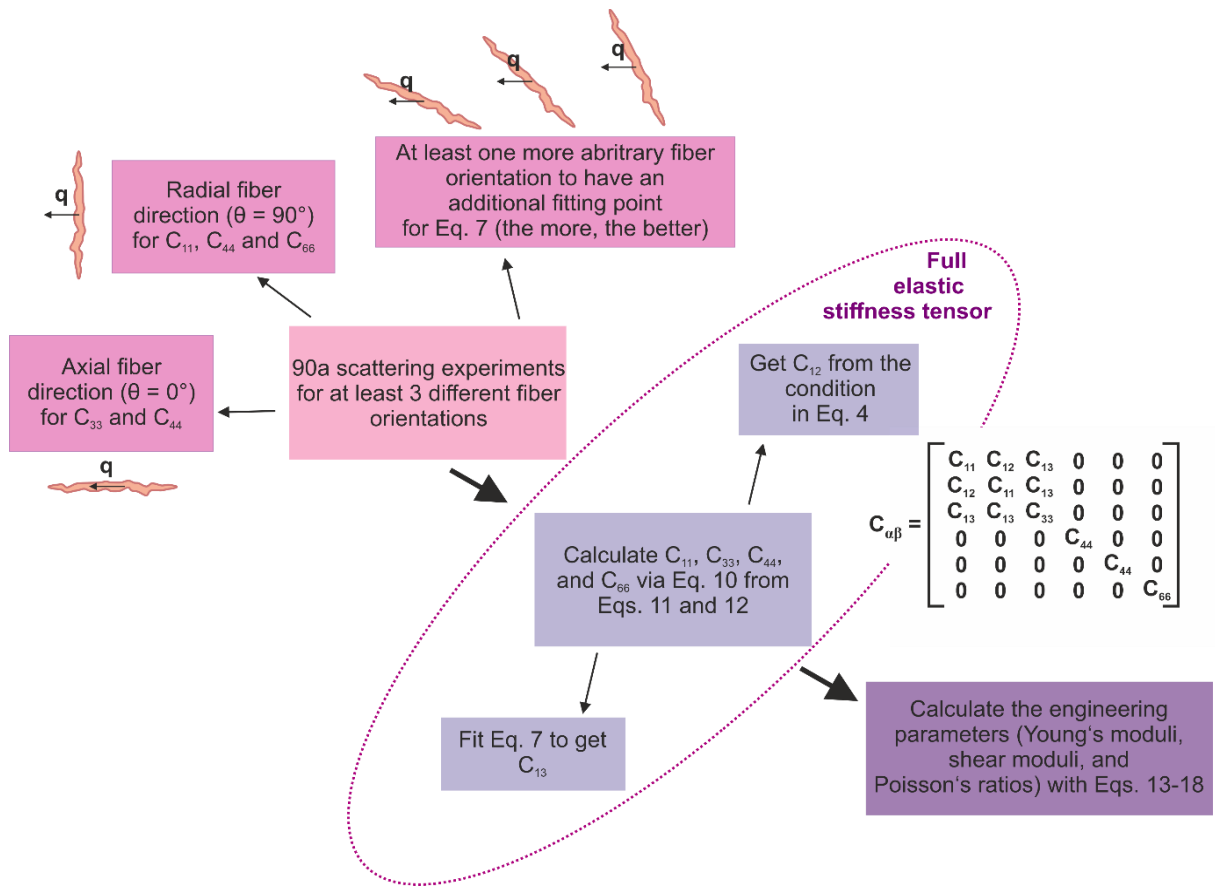

*Figure 3: Summary of the basic experimental and theoretical steps to obtain the elastic stiffness tensor of a fiber. It should be noted that this is an idealized scheme which assumes that all the diagonal components of the elastic stiffness tensor in Eq. 3 are obtained from a minimum number of experiments.*

**Results**

Assuming a hexagonal symmetry and using the Christoffel's solutions for this symmetry (Eq. 7 – 9), the elastic stiffness tensor of a homogeneous, cellulosic viscose fiber was investigated with BLS in different fiber orientations.

First, BLS measurements of the viscose fiber were performed in the axial direction of the fiber (fiber orientation $\theta = 0°$). As described in Section 2.5, in axial fiber direction, the two quasi-transverse waves (T1 and T2) degenerate and are equal ($T1 = T2 = T$). To differ between the quasi-longitudinal (L) wave and the quasi-transverse (T) wave, polarizers are inserted into the beam path and depending on their alignment either L or T are vanishing or at least strongly reduced. In parallel-polarizer alignment (HH – incident light horizontally polarized and scattered light horizontally polarized), T is not visible, whereas in cross-polarizer alignment (HV), L should vanish. The HH and HV spectra are the top two spectra presented in Figure 4a. Here, it is surprising to see in HV three different peaks. In HH, only two peaks show up. Therefore, the peak at the lowest frequency shift is clearly corresponding to a quasi-transverse wave (T). However, the other two peaks appear in both polarizer alignments.

BLS measurements were continued at fiber orientations $\theta = 25°, 45°, 65°$, and $90°$, going from axial to radial fiber direction, as presented in Fig. 4a. For $\theta = 25°$ and $\theta = 45°$, the trend is the same as for $\theta = 0°$. Three peaks are visible. However, at $\theta = 65°$ only two peaks are still visible and at $\theta = 90°$ only one peak at the highest frequency shift is seen. With increasing orientation angle, the signal to noise level increased, especially at lower frequency shift values. Therefore, only filtering of the data (not depicted here) made it possible to obtain one additional peak for the 90° orientation. Interestingly, at $\theta = 90°$, the peak at high frequency shift values coincides with the backscattering peak (180a scattering geometry). This indicates that the measured peak at frequency shift values of around 20 $GHz$ is the result of a spontaneous backscattering process and, therefore, the peak at about 13 $GHz$ for the axial fiber direction, which is decreasing with increasing fiber angle, corresponds to the quasi-longitudinal (L) wave of the fiber. For the quasi-transverse waves, only one wave (T) is found throughout all fiber orientations, indicating that $T1$ and $T2$ are in the same range. Furthermore, at fiber orientations $\theta > 45°$, it is not possible to distinguish T from the noise anymore.

In Fig. 4b, the frequency shifts of 90a and 180a scattering obtained from the spectra in Fig. 4a with both polarizer alignments HH and HV are plotted. From the BLS spectra of the fiber, the frequency shifts of the quasi-longitudinal (L) wave and the quasi-transverse (T) wave at all fiber orientations $\theta$ are determined. As a consequence of the quasi-transverse waves being equal ($T1 = T2$): $C_{44} = C_{66}$, which reduces the independent stiffness tensor components further down to four: $C_{11}, C_{33}, C_{44}$, and $C_{13}$ ($C_{12} = C_{11} - 2C_{44}$.). All the diagonal stiffness tensor components are directly obtained from the BLS measurements via Eqs. 11 and 12 whereas $C_{13}$ is determined by solving the Christoffel equation's solution (Eq. 7).

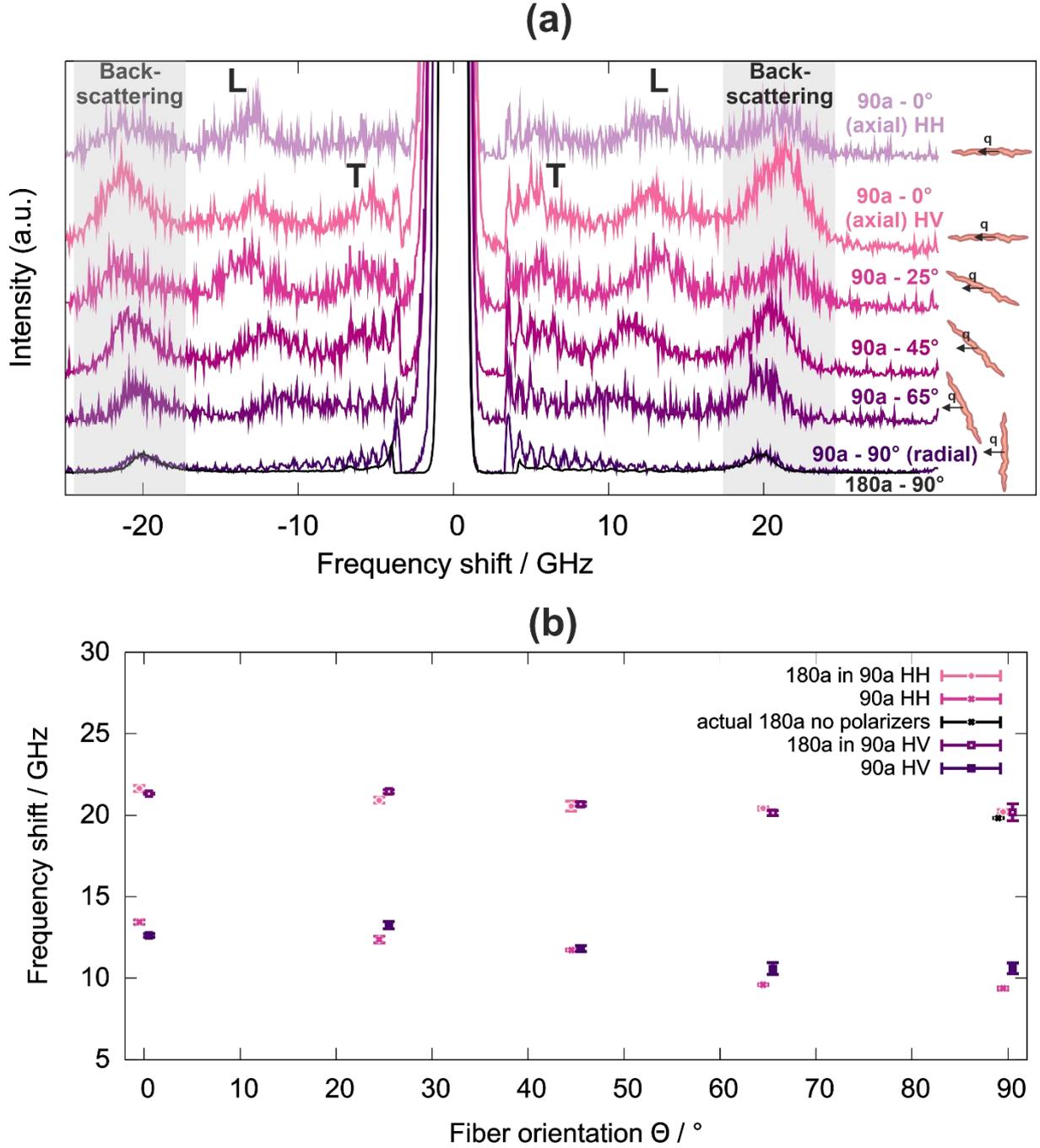

*Figure 4: (a) Spectra for all fiber orientations in the 90a scattering geometry (raw data). A quasi-longitudinal L and a quasi-transverse T peak are visible and with increasing rotation angle (going from axial fiber alignment to radial fiber alignment) the peaks are moving to lower frequency shifts. The peak in the greyed-out area is coinciding at fiber orientation $\theta = 90°$ with the longitudinal peak of the 180a scattering geometry. (b) Quasi-longitudinal frequency shifts of the (a) 90a and (b)180a scattering peaks of both polarizer orientations as a function of fiber orientation.*

In Fig. 5, the fitting of the Christoffel equation solutions (Eq. 7 – 9) obtained for the viscose fiber measured in the BLS experiments of this work is presented for the quasi-longitudinal (L) and the quasi-transverse (T1 and T2) waves together with the experimental results. In the axial fiber direction, at $\theta = 0°$, L exhibits the largest sound velocity $V$ of around $5000 \, m/s$ (corresponding to a frequency shift of about 13 $GHz$). This value decreases with increasing fiber orientation (towards $\theta = 90°$) to about $V = 3500\ m/s$ (9 $GHz$ frequency shift). Since T1 and T2 are considered equal, they show only slight variation of their curves and stay quite constant over all fiber orientations at a sound velocity $V = 2000\ m/s$ ($\Delta f = 5.3\ GHz$ frequency shift).

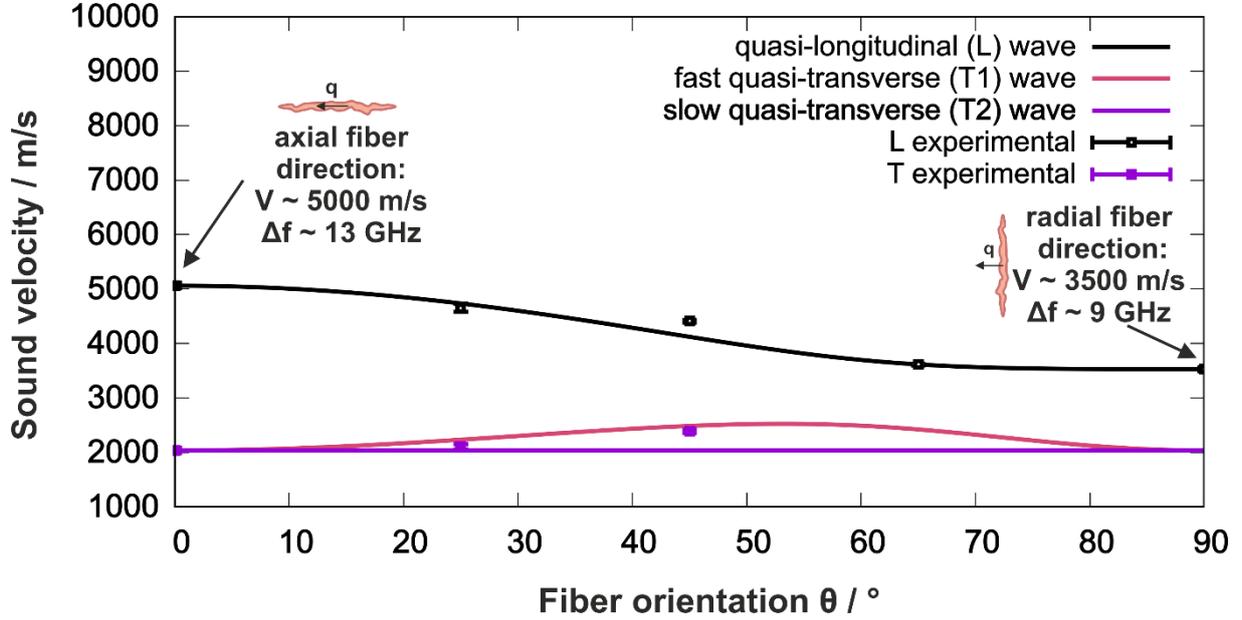

*Figure 5: Overview of the Christoffel's solution fitting results obtained for the viscose fiber together with the experimental results.*

Finally, in Eq. 19, the elastic stiffness tensor of the viscose fiber is depicted with $C_{11} = 18.6\ GPa$, $C_{33} = 38.4\ GPa$, $C_{44} = C_{66} = 6.2\ GPa$, $C_{12} = 6.2\ GPa$ and $C_{13} = 6.6\ GPa$:

$$C_{\alpha\beta} = \begin{bmatrix} 18.6 & 6.2 & 6.6 & 0 & 0 & 0 \\ 6.2 & 18.6 & 6.6 & 0 & 0 & 0 \\ 6.6 & 6.6 & 38.4 & 0 & 0 & 0 \\ 0 & 0 & 0 & 6.2 & 0 & 0 \\ 0 & 0 & 0 & 0 & 6.2 & 0 \\ 0 & 0 & 0 & 0 & 0 & 6.2 \end{bmatrix}.$$  (Equation 19)

With the knowledge of the full elastic stiffness tensor, one can employ Eq. 13 – 18 to determine the engineering mechanical parameters: axial Young's modulus $E_{axial}$, radial Young's modulus $E_{radial}$, shear moduli $G_{13} = G_{12}$, and the Poisson's ratios $\sigma_{13}$ and $\sigma_{12}$. They are summarized in Table 1 and compared to parameters obtained from classical micromechanical testing (tensile testing, nanoindentation and torsion testing).

*Table 1: Summary of the viscose fiber's engineering mechanical parameters (Young's moduli in axial and radial fiber direction, shear moduli, and Poisson's ratios) based on the BLS measurements obtained from Eq. 13 – 18 as well as determined by mechanical testing in this work and in literature.*

|  | Axial Young's modulus $E_{axial}$ / GPa | Radial Young's modulus $E_{radial}$ / GPa | Shear moduli $G_{13} = G_{12}$ / GPa | Poisson's ratio $\sigma_{13}$ / - | Poisson's ratio $\sigma_{12}$ / - |
|---|---|---|---|---|---|
| BLS from this work | 34.9 | 16.0 | 6.2 | 0.27 | 0.29 |
| Mechanical testing | 7.2 ± 1.3[*] | 4-5[**] | 3-4[***] | - | 0.25[****] |

[*] Tensile testing from this work; [**] nanoindentation on the same viscose fiber type [6] only a so-called reduced modulus is obtained; [***] Torsion testing of wood-based fibers [12]; [****] nanoindentation of wood-based fibers [39]

**Discussion**

The determination of the elastic stiffness tensor of a viscose fiber with the 90a scattering geometry approach was demonstrated. At different fiber orientations $\theta$, a difference in the quasi-longitudinal and quasi-transverse wave peaks was found. It was shown that the backscattering peak is also measured during these measurements. Such backscattering peaks were first observed in thin samples [40] and also seem to be a trait of fibrous materials, e. g., marine sponge spicules and glass fibers [25]. Interestingly, we also find that the viscose fibers exhibit optical waveguide properties [41] which result in the observed spontaneous Brillouin backscattering. As presented in Fig. 4, there is a peak at about $(20 - 21)$ $GHz$ in all spectra which would lead to a sound velocity of $(7600 - 8100)$ $m/s$ if evaluated with the 90a scattering geometry equation (Eq. 10). Such high values for the sound velocity seem unrealistic given that in the literature, ultrasonic experiments on wood, which have a higher cellulose crystallinity than the investigated viscose fibers, reported a sound velocity of only $(5000 - 6000)$ $m/s$ in the axial fiber direction [42]. Viscose fibers are, however, also not fully crystalline ($< 50$ % crystallinity [1]); the microstructure consists of crystalline and amorphous regions and with our BLS measurements, we are most likely averaging over an area of a few micrometers [43,44]. Furthermore, in Fig. 4a, it is clearly visible that the 180a and 90a spectra of the fiber oriented in radial direction are overlapping which strengthens the argument of the backscattering peak.

Calculating the sound velocity for the 180a scattering geometry with a refractive index $n$ of 1.34, as measured for the same viscose fiber type in [27] with holographic phase microscopy, results in sound velocities between $(4300 - 4700)$ $m/s$. However, this value may be too low given also that viscose fibers themselves are highly birefringent (1.550 in axial and 1.515 in radial fiber direction, respectively [45]). Evidence of the birefringence is also visible in our results in Fig. 4b. Here, the frequency shifts of the quasi-longitudinal peaks of the 90a and 180a scattering show a difference at all fiber orientations between polarizer alignment HH and HV. Therefore, using the 90a scattering geometry for fibrous materials is advantageous because the refractive index is not needed for the evaluation of the stiffness components.

In Fig. 5, the Christoffel's solution fitting results indicate the expected trend for an anisotropic fiber. In axial direction, fiber orientation $\theta = 0°$, the sound velocity exhibits higher values compared to the radial direction ($\theta = 90°$) for the quasi-longitudinal wave. In the Brillouin spectra of the viscose fiber (Fig. 4a) only one quasi-transverse peak was found. This resulted in an additional constraint of $C_{44} = C_{66}$ and a further reduction of the independent stiffness tensor components to four: $C_{11}$, $C_{33}$, $C_{13}$, and $C_{44}$, with $C_{12} = C_{11} - 2C_{44}$.

It should be noted that $C_{44} = C_{66}$ is an assumption which is made because no quasi-transverse wave (T2) was observed in any of the BLS measurements. This might indicate that the difference between $C_{44}$ and $C_{66}$ is too small to resolve. Determining the elastic stiffness constant $C_{66}$ requires the knowledge of the slow quasi-transverse wave T2 according to Eq. 9. However, we only observed fast quasi-transverse waves through the entire set of measurements. Determining whether T1 and T2 are overlapping or if the slow quasi-transverse peaks are too close to the Rayleigh peak in the obtained spectra and, therefore, hidden in the noise is challenging. Furthermore, according to Eq. 12 we expect $C_{66} \leq C_{44}$ as $C_{66}$ is connected to the slow and $C_{44}$ to the fast quasi-transverse wave at this fiber orientation, unless the transverse waves show an intersection point [38]. Consequently, we have to make an assumption to determine the constant $C_{66}$.

With the elastic stiffness tensor fully described in Eq. 19, the engineering mechanical parameters were calculated and are listed in Table 1. The Young's moduli in the axial and radial fiber direction are $34.9\ GPa$ and $16.0\ GPa$, respectively. These values are higher than what is obtained by micromechanical testing of viscose fibers. The investigated viscose fibers can withstand only small stresses ($< 100\ MPa$) and have a quite high breaking strain ($\sim 10\ \%$) in comparison to other natural cellulosic fibers. Their Young's modulus determined by tensile experiments has a mean value of $7.3\ GPa$. In the radial fiber direction, atomic force microscopy-based nanoindentation for the same fiber type gives a reduced modulus of about $5\ GPa$ [6]. $E_{axial}$ obtained by BLS is higher by a factor of five compared to Young's modulus from tensile testing and $E_{radial}$ by a factor of three compared to the reduced modulus from nanoindentation measurements. The axial shear modulus $G_{LT} = C_{44}$ is the relevant material parameter for twisting the fiber about its axial direction [12], while the in-plane shear modulus $G_{TT} = C_{66}$ is the relevant material parameter for shearing the fiber in radial direction (i.e., like a bolt loaded in shear). Since the shear moduli are directly determined from the $C_{44}$ and $C_{66}$ values (Eqs. 15 and 16), and we concluded that $C_{44} = C_{66}$, both shear moduli are equal with a value of $6.2\ GPa$. This is roughly a factor of two higher than what was found in the literature in torsional tests for wood-based fibers [12], which have a smaller cross-sectional area than the investigated viscose fiber. In summary, the results for the Young's and shear moduli from (quasistatic) mechanical testing are lower by a factor 2 to 5. This is plausible considering that BLS probes at much higher frequencies, which for viscoelastic materials like cellulosic fibers will result in a higher stiffness [13]. Still, by using loading rate dependencies obtained by tensile testing from the literature [13], the increase due to viscoelastic effects accounts at best for about a factor of two increase in axial direction if one assumes a linear relationship over all frequency regimes. The discrepancy may however be accounted for by additional fast relaxation processes between the two frequency regimes. For both Poisson's ratios $v_{13}$ and $v_{12}$ similar values of 0.27 and 0.29, respectively, were found. These values are in relatively good agreement with measured and assumed literature values on wood-based cellulosic fibers [39,46].

**Conclusions**

In this work, Brillouin light scattering spectroscopy (BLS) was used to determine the full elastic stiffness tensor of a man-made cellulose fiber – a so-called viscose fiber. The experimental and theoretical procedure presented can be applied to any fibrous material that is transversely isotropic, i.e., with different stiffness properties in axial and radial direction. The viscose fiber was measured in the 90a scattering geometry at five different fiber orientations – $\theta = 0°$ (axial fiber direction), 25°, 45°, 65°, and 90° (radial fiber direction) – which enabled the determination of all stiffness tensor components directly from the BLS measurements. The 90a scattering geometry is not often applied to biological matter, however, it is very advantageous because – compared to the more commonly applied backscattering geometry (180a) – the refractive index is not needed to calculate the sound velocity. In this work, it was further demonstrated that – for materials with optical waveguide properties – the 90a scattering spectra can also contain 180a scattering information. This way, one scattering setup allows for the measurement of two different scattering processes. Namely a single spectral measurement can in principle be used to determine the entire stiffness tensor. Furthermore, the optical waveguiding modes could be promising for future applications in biomaterial-based photonic circuits.

For the viscose fiber, the quasi-longitudinal wave in axial fiber direction resulted in sound velocity values, which agreed well with the literature values on wood tested by ultrasonic methods. Only one quasi-transverse wave was found which resulted in the assumption $C_{44} = C_{66}$. Overall, the elastic stiffness tensor is assumed to be composed of four independent components: $C_{11}$, $C_{33}$, $C_{44}$, and $C_{13}$. By their determination, it was possible to calculate engineering-relevant material parameters such as Young's moduli, shear moduli, and Poisson's ratios. The Young's moduli in axial and radial fiber direction and the shear moduli were lower by factors of 2 to 5 compared to results from micromechanical testing (tensile testing, nanoindentation, torsional shear testing). This is not surprising, considering that cellulosic fibers are inherently viscoelastic, which leads to an increase in the measured stiffness with increasing deformation rate. Since BLS probes the material at a higher frequency regime (GHz), it is clear that this corresponds to a much higher strain rate than the (quasi-static) mechanical tests (Hz). Accordingly, the stiffness values measured by BLS are considerably higher. The Poisson's ratios were in a similar range compared to previous literature. Overall, the proposed approach provides *one consistent* method to measure all stiffness properties of transversely isotropic fibers. This consistent approach provides more comprehensive and distinct results than the collection of different micromechanical testing methods required for testing the Young's moduli, shear moduli and Poisson's ratios in axial and radial direction of a fiber.

**Acknowledgements**

C. C. acknowledges the Hertha Firnberg program of the Austrian Science Fund (FWF) for funding. This research was funded in whole by the Austrian Science Fund (FWF) [T 1314-N]. For open access purposes, the author has applied a CC BY public copyright license to any author accepted manuscript version arising from this submission. Special thanks to Angela Wolfbauer for sample preparation. Lots of gratitude to John R. Sandercock from the Table Stable Ltd. for providing the possibility to perform measurements at his company.

**Conflict of Interests**

The authors declare that they have no known competing financial interests or personal relationships that could have appeared to influence the work reported in this article.

**Data Access statement**

The experimental data of this work is available from the corresponding author upon request.

**Ethics statement**

The authors declare that no ethical issues arise in this work.